\documentclass[aps,rpb,twocolumn]{revtex4}
\usepackage{graphics,epsfig,amsmath,lscape,colordvi}

\begin{document}

\title{Calculations of Optical Properties in Strongly Correlated Materials}
\author{V. S. Oudovenko$^{\dagger,*}$, G. P\'alsson$^{*}$, S. Y. Savrasov$%
^{\ddagger}$, K. Haule$^{*}$ and G. Kotliar$^{*}$ }
\affiliation{$^{\dagger}$Bogoliubov Laboratory for Theoretical Physics, %
Joint Institute for Nuclear Research, 141980 Dubna, Russia\\
$^{*}$Center for Materials Theory, Department of Physics and Astronomy,
Rutgers University, Piscataway, NJ 08854\\
$^{\ddagger}$Department of Physics, New Jersey Institute of Technology,
Newark, NJ 07102}
\date{\today}

\begin{abstract}
We present a new method to calculate optical properties of strongly
correlated systems. It is based on dynamical mean-field theory and it uses
as an input realistic electronic structure obtained by local density
functional calculations. Numerically tractable equations for optical
conductivity, which show a correct non-interacting limit, are derived.
Illustration of the method is given by computing optical properties of the
doped Mott insulator La$_{1-x}$Sr$_x$TiO$_3$.
\end{abstract}

\maketitle

\draft

\section{Introduction}

Optical spectral functions such as conductivity or reflectivity are very
important characteristics of solids which give us a direct probe of their
electronic structure. In the past, very powerful numerical techniques \cite%
{DFTbook} based on density functional theory (DFT)\ and local density
approximation (LDA) have been developed, which allowed to access the
one--electron spectrum in real materials via association of LDA energy bands
with the real excitation energies. This approach works well for weakly
correlated systems, where, for example, optical properties can be directly
computed \cite{Maks} via the knowledge of the band structure and the dipole
matrix elements of the material. Furthermore, for weakly correlated
materials LDA is a good starting point for adding perturbative corrections
in the screened Coulomb interactions following the GW approach \cite%
{gwreview}.

Unfortunately, the treatment of materials with strong electronic
correlations is not possible within this framework. Strong on--site Coulomb
repulsion modifies the one--electron spectrum via appearance of satellites,
Hubbard bands, strongly renormalized Kondo--like states, etc., which are no
longer obtainable using static mean--field theories such as Hartree--Fock
theory or LDA. The wave functions in strongly correlated systems are not
representable by single--Slater determinants and dynamical self--energy
effects become important, thus requiring a new theoretical treatment based
on the dynamical mean--field theory (DMFT) ~\cite{DMFTreview}. Recent
advances ~\cite{Anisimov:1997} in merging the DMFT with realistic LDA based
electronic structure calculations have already led to solving such long
standing problems as, e.g., temperature dependent magnetism of Fe and Ni
\cite{Licht}, volume collapse in Ce \cite{CeVolume}, and huge volume
expansion of Pu~\cite{Nature}.

In the present work we develop a new approach which allows us to
calculate the optical properties of strongly correlated materials
within the combined LDA and DMFT framework. We discuss the
expressions for optical conductivity using self--energies and
local Green's functions, which are numerically tractable and
correctly reproduce the limit of non--interacting electrons. We
also check the limit of strong correlations by applying the
method to three--band Hubbard Hamiltonian. Results of this test
reproduce the available experimental and theoretical data with
very good accuracy. We demonstrate the applicability of the
present scheme on the example of doped Mott insulator
La$_{1-x}$Sr$_{x}$ TiO$_{3}$, where we compare the results of our
new calculations with the LDA predictions and experiment.

The paper is organized as follows. In the next section \ref{sec:method} we
describe the method for calculation of the optical conductivity. Application
of the method to doped La$_{1-x}$Sr$_{x}$TiO$_{3}$ is described and analyzed
in section \ref{sec:results} which is followed by conclusions presented in section~\ref%
{sec:conclusion}. Some technical details of the calculations and the
downfolding and upfolding procedures are given in Appendix ~\ref{sec:app_se}.

\section{Method}

\label{sec:method}

To calculate the optical response functions we utilize the dynamical mean
field approach where the self--energy of the many--body problem is
approximated by a local operator $\Sigma (\omega )$ which is, however,
frequency dependent. A physical transparent description of this method can
be achieved by introducing an interacting analog of Kohn--Sham particles, $%
\psi _{{\mathbf{k}}j}(\mathbf{r},\omega )\equiv \psi _{{\mathbf{k}}j\omega }$%
, which reproduce the local portion of the Green's function in a similar way
as the non--interacting Kohn--Sham particles $\psi _{{\mathbf{k}}j}(\mathbf{r%
})$ reproduce the density of the solid in its ground state. This spectral
density functional approach \cite{Functional} has an advantage that the
\textbf{k}--integrated excitation properties (such, e.g., densities of
states) can now be associated with the real one--electron spectra. The
optical transitions between the interacting quasiparticles $\psi _{{\mathbf{k%
}}j\omega }$ allow the excitations between incoherent and coherent parts of
the spectra (e.g., between Hubbard and quasiparticle bands) which are
intrinsically missing in static mean--field approaches such as DFT but are
present in real strongly correlated situations.

In order to find the quasiparticles living at a given frequency $\omega $ we
solve the Dyson equation with the LDA potential $V_{eff}$ and the frequency
dependent correction ${\Sigma }(\omega )-{\Sigma }_{dc}$, i.e.
\begin{equation}
(-\nabla ^{2}+V_{eff}+{\Sigma }(\omega )-{\Sigma }_{dc}-\epsilon _{{\mathbf{k%
}}j\omega })\psi _{{\mathbf{k}}j\omega }^{R}=0.  \label{Dyson}
\end{equation}%
A double counting term ${\Sigma }_{dc}$ appears here to account
for the fact that $V_{eff}$ is the average field which acts on
both heavy (localized) and light (itinerant) electrons. Note,
that due to non--Hermitian nature of the problem, both ``right"
$\psi ^{R}$and ``left" $\psi ^{L}$ eigenvectors should be
considered, the latter being the solution of the same Dyson
equation (\ref{Dyson}) with $\psi $ placed on the left. The local
Green's function is constructed from the eigenvectors and
eigenvalues in the following way
\begin{equation}
G(\omega )=\sum_{{\mathbf{k}}j}\frac{\psi _{{\mathbf{k}}j\omega }^{R}\psi _{{%
\mathbf{k}}j\omega }^{L}}{\omega +\mu -\epsilon _{{\mathbf{k}}j\omega }}.
\label{G_loc}
\end{equation}%
The local self--energy is calculated from the corresponding impurity problem
which is defined by the DMFT self--consistency condition
\begin{equation}
G(\omega )=\left( \omega -E_{imp}-\Sigma (\omega )-\Delta _{imp}(\omega
)\right) ^{-1},  \label{G_imp}
\end{equation}%
where $\Delta _{imp}$ is the impurity hybridization matrix and $E_{imp}$ are
the impurity levels. From known $\Delta _{imp}(\omega )$, $E_{imp}$ and
Coulomb interaction $U,$ the solution of the Anderson impurity problem then
delivers the local self--energy ${\Sigma }(\omega )$. The system of
equations (\ref{Dyson}), (\ref{G_loc}) and (\ref{G_imp}), together with an
impurity solver, i.e., a functional $\Sigma \lbrack \Delta _{imp}(\omega
),E_{imp},U]$, is thus closed.

Solution of the Anderson impurity model can be carried out by available
many--body technique \cite{DMFTreview} such as the Quantum Monte Carlo (QMC)
method \cite{QMC} which will be used in our work. In practice \cite%
{Anisimov:1997,Nature}, we utilize the LDA+DMFT approximation and treat only
the $d$--electrons of Ti as strongly correlated thus requiring full energy
resolution. All other electrons are assumed to be well described by the LDA.
The Dyson equation is solved on the Matsubara axis for a finite set of
imaginary frequencies $i\omega _{n}$ using a localized orbital
representation such, e.g., as linear muffin-tin orbitals (LMTOs) \cite%
{Andersen:1975} for the eigenvectors $\psi _{\mathbf{k}j\omega }$.

The optical conductivity can be expressed via equilibrium state
current--current correlation function \cite{Mahan} and is given by:
\begin{eqnarray}
\sigma _{\mu \nu }(\omega ) &=&\pi e^{2}\int\limits_{-\infty }^{+\infty
}d\varepsilon \phi _{\mu \nu }(\varepsilon +\omega /2,\varepsilon -\omega
/2)\times  \notag  \label{sigma} \\
&&\frac{f({\varepsilon -\omega /2)-}f({\varepsilon +\omega /2)}}{\omega },
\end{eqnarray}%
where $e$ is free electron charge, $f(\varepsilon )$ is the Fermi function
and the transport function $\phi _{\mu \nu }(\varepsilon ,\varepsilon
^{\prime })$ is defined as
\begin{equation}
\phi _{\mu \nu }(\varepsilon ,\varepsilon ^{\prime })=\frac{1}{\mathcal{V}}%
\sum\limits_{\mathbf{k}jj^{\prime }}{}Tr\left\{ \nabla _{\mu }{\hat{\rho}_{%
\mathbf{k}j}(\varepsilon )\nabla _{\nu }{\rho }_{\mathbf{k}j^{\prime
}}(\varepsilon ^{\prime })}\right\} ,  \label{transp}
\end{equation}%
with $\mathcal{V}$ being the unit cell volume and
\begin{equation}
{\rho }_{\mathbf{k}j}(\varepsilon )=-\frac{1}{2\pi i }\left( {G}_{\mathbf{k}%
j}(\varepsilon )-{G}_{\mathbf{k}j}^{\dagger }(\varepsilon )\right) ,
\label{rho}
\end{equation}%
is expressed via retarded one--particle Green's function, ${G}_{\mathbf{k}
j}(\varepsilon )$. Using the solutions $\epsilon _{\mathbf{k}j\omega }$ and $%
\psi _{\mathbf{k}j\omega }$ of the Dyson equation (\ref{Dyson}) we express
the optical conductivity in the form:
\begin{eqnarray}
&&\sigma _{\mu \nu }(\omega )=-\frac{e^{2}}{4\pi }\sum_{ss^{\prime }=\pm
1}ss^{\prime }\sum_{\mathbf{k}jj^{\prime }}\int\limits_{-\infty }^{+\infty
}d\varepsilon \frac{M_{\mathbf{k}jj^{\prime }}^{ss^{\prime },\mu \nu
}(\varepsilon ^{-},\varepsilon ^{+})}{\omega +\epsilon _{\mathbf{k}%
j\varepsilon ^{-}}^{s}-\epsilon _{\mathbf{k}j^{\prime }\varepsilon
^{+}}^{s^{\prime }}}\times  \notag  \label{sigma2} \\
&&\left[ \frac{1}{\varepsilon ^{-}+\mu -\epsilon _{\mathbf{k}j\varepsilon
^{-}}^{s}}{-}\frac{1}{\varepsilon ^{+}+\mu -\epsilon _{\mathbf{k}j^{\prime
}\varepsilon ^{+}}^{s^{\prime }}}\right] \frac{f(\varepsilon
^{-})-f(\varepsilon ^{+})}{\omega },
\end{eqnarray}%
where we have denoted $\varepsilon ^{\pm }=\varepsilon \pm \omega /2$, and
used the shortcut notations $\epsilon _{\mathbf{k}j\varepsilon }^{+}\equiv
\epsilon _{\mathbf{k}j\varepsilon }$, $\epsilon _{\mathbf{k}j\varepsilon
}^{-}=\epsilon _{\mathbf{k}j\varepsilon }^{\ast }$.

The matrix elements $M_{\mathbf{k}jj^{\prime }}$ are generalizations of the
standard dipole allowed transition probabilities which are now defined with
the right and left solutions $\psi ^{R}$and $\psi ^{L}$ of the Dyson
equation:
\begin{eqnarray}
&&M_{\mathbf{k}jj^{\prime }}^{ss^{\prime },\mu \nu }(\varepsilon
,\varepsilon ^{\prime })=  \label{matel} \\
&&\int (\psi _{\mathbf{k}j\varepsilon }^{s})^{s}\nabla _{\mu }(\psi _{%
\mathbf{k}j\varepsilon ^{\prime }}^{-s^{\prime }})^{s^{\prime }}d\mathbf{r}%
\int (\psi _{\mathbf{k}j^{\prime }\varepsilon ^{\prime }}^{s^{\prime
}})^{s^{\prime }}\nabla _{\nu }(\psi _{\mathbf{k}j\varepsilon }^{-s})^{s}d%
\mathbf{r},  \notag
\end{eqnarray}
where we denoted $\psi _{\mathbf{k}j\varepsilon }^{+}=\psi _{\mathbf{k}
j\varepsilon }^{L}$ , $\psi _{\mathbf{k}j\varepsilon }^{-}=\psi _{\mathbf{k}
j\varepsilon }^{R}$ and assumed that $(\psi _{\mathbf{k}j\varepsilon
}^{s})^{+}\equiv \psi _{\mathbf{k}j\varepsilon }^{s}$ and $(\psi _{\mathbf{k}
j\varepsilon }^{s})^{-}=\psi _{\mathbf{k}j\varepsilon }^{s\ast }$.
Expressions (\ref{sigma2}), (\ref{matel}) represent generalization of the
optical conductivity formula for the case of strongly correlated systems,
and involve the extra internal frequency integral appearing in Eq.~(\ref%
{sigma2} ).

Let us consider the non-interacting limit when ${\Sigma }(\omega )-{\Sigma }
_{dc}\rightarrow i\gamma \rightarrow 0.$ In this case, the eigenvalues $%
\epsilon _{\mathbf{k}j\varepsilon }=\epsilon _{\mathbf{k}j}+i\gamma ,\psi _{%
\mathbf{k}j\varepsilon }^{R}\equiv |\mathbf{k}j\rangle ,$ $\psi _{\mathbf{k}
j\varepsilon }^{L}\equiv |\mathbf{k}j\rangle ^{\ast }\equiv \langle \mathbf{k%
}j|$ and the matrix elements $M_{\mathbf{k}jj^{\prime }}^{ss^{\prime },\mu
\nu }(\varepsilon ,\varepsilon ^{\prime })$ are all expressed via the
standard dipole transitions $|\langle \mathbf{k}j|\nabla |\mathbf{k}
j^{\prime }\rangle |^{2}.$ Working out the energy denominators in the
expression (\ref{sigma2}) in the limit $i\gamma \rightarrow 0$ and for $%
\omega \neq 0$ leads us to the usual form for the conductivity which for its
interband contribution can be written as:
\begin{eqnarray}
\sigma _{\mu \nu }(\omega ) &=&\frac{\pi e^{2}}{\omega }\sum_{\mathbf{k}%
,j^{\prime }\neq j}\langle \mathbf{k}j|\nabla _{\mu }|\mathbf{k}j^{\prime
}\rangle \langle \mathbf{k}j^{\prime }|\nabla _{\nu }|\mathbf{k}j\rangle
\times  \notag  \label{sigmani} \\
&&[f(\epsilon _{\mathbf{k}j})-f(\epsilon _{\mathbf{k}j^{\prime }})]\delta
(\epsilon _{\mathbf{k}j}-\epsilon _{\mathbf{k}j^{\prime }}+\omega ).
\end{eqnarray}

To evaluate the expression $\sigma _{\mu \nu }(\omega )$ in Eq.~ (\ref%
{sigma2}) numerically, we need to perform integration over $\varepsilon $
and pay a special attention to the energy denominator $1/(\omega +\epsilon _{%
\mathbf{k}j\varepsilon ^{-}}^{s}-\epsilon _{\mathbf{k}j^{\prime
}\varepsilon ^{+}}^{s^{\prime }})$. To calculate the integral
over $\varepsilon $ we divide frequency domain into discrete set
of points $\varepsilon _{i}$ and assume that the eigenvalues
$\epsilon _{\mathbf{k}j\varepsilon }$ and eigenvectors $\psi
_{\mathbf{k}j\varepsilon }$ to zeroth order can be approximated
by their values at the middle between each pair of points. In
this way,\ the integral is replaced by the discrete sum over internal grid $%
\varepsilon _{i}$ defined for each frequency $\omega .$ To deal with the
strong momentum dependence of $1/(\omega +\epsilon _{\mathbf{k}j\varepsilon
^{-}}^{s}-\epsilon _{\mathbf{k}j^{\prime }\varepsilon ^{+}}^{s^{\prime }})$,
linearization of the denominator with respect to $\mathbf{k}$ should be
performed as it is done in the tetrahedron method of Lambin and Vigneron
\cite{Lambin}. On the other hand, the difference between single poles
(expression in square brackets of Eq.(\ref{sigma2})), after integration over
frequency becomes a smooth function of $\mathbf{k}$ and can be treated
together with the current matrix elements, i.e by linearizing the numerator.
The described procedure produces a fast and accurate algorithm for
evaluating the optical response functions of a strongly correlated material.

\section{Application of the method}

\label{sec:results}

To illustrate the method of the optical conductivity calculation in a
strongly correlated system we chose paramagnetic doped Mott insulator La$%
_{1-x}$Sr$_{x}$TiO$_{3}$. LDA cannot reproduce insulating
behavior of this system already at $x=0,$ which emphasizes the
importance of correlation effects. Upon doping the system becomes
a correlated metal, which at $x=1$ (SrTiO$_{3}$) should be
considered as a standard band insulator. Photoemission
experiments \cite{Photoemis} as a function of doping display both
a lower Hubbard band located at near energies 2~eV below the Fermi
level $E_{F\text{ }}$ and a quasiparticle band centered at
$E_{F}$. Previous DMFT based calculations
\cite{Anisimov:1997,AnisLTO} of the density of states used
$t_{2g}$ degenerate bands of Ti found near $E_{F}$ and reproduced
both these features with a good accuracy. The studies of the
optical properties for LaTiO$_{3}$ with the less accurate LDA+U\ method \cite%
{LDAU} have been also carried out \cite{Solovyev}.

We have calculated the electronic structure of La$_{1-x}$Sr$_{x}$TiO$_{3}$
using the LDA+DMFT method. A cubic crystal structure with 5 atoms per unit
cell is utilized which is a simplified version of a fully distorted 20
atoms/cell superlattice. Since the self--energy effects are crucial for the
states near the Fermi energy, we treat correlations only on the downfolded~$%
t_{2g}$ orbitals of Ti atoms as suggested previously \cite%
{Anisimov:1997,AnisLTO}. The Anderson impurity model is solved using Quantum
Monte Carlo method with Hubbard parameter $U=6$~eV at $T=1/\beta =1/32$ of
Ti $t_{2g}$ bandwidth which delivers the self--energy ${\Sigma }(\omega )$
for these orbitals using the self--consistent DMFT framework. The
applicability of QMC is justified since temperature in our simulation is
well below the coherence energy, which is about 1/8 of the bandwidth. We
also limit our consideration by dopings $x$ larger than 10~per cent to stay
below the coherence temperature. Once the self--energy is obtained, we
upfold it back into the full orbital space which delivers the one--electron
spectrum of the system with correlation effects taken into account. Detailed
description of downfolding/upfolding procedures to get the self--energy is
given in Appendix ~\ref{sec:app_se}.

To treat doping away from $x=0$ the self--energy is allowed to
change self--consistently while the one--electron Hamiltonian is
assumed to be independent on doping. We then evaluate the
frequency--dependent eigenvalues $\epsilon _{\mathbf{k}j\omega
},\psi _{\mathbf{k}j\omega }$ as functions of doping. This allows
us to evaluate the energy and doping dependent optical
conductivity integrals both in $\mathbf{k}$- and $\epsilon
$-spaces. The integrals over momentum are taken on the
$(10,10,10)$ mesh using the tetrahedron method of
Ref.~\cite{Lambin}. To check the convergence we also performed
the calculations on the (6,6,6) mesh which produces the
conductivity within 5~per cent of accuracy. The energy integration
mesh was chosen to have a step equal to 0.01~eV. We also broaden
the imaginary part of the self--energy for non--interacting bands
with 0.0004~eV. This reproduces the LDA density of states of the
studied compound within the accuracy of 1--2 per cent.

We first discuss the undoped case with $x=0$ which corresponds to the
insulator with a small gap equal to 0.2--0.5~eV. Model calculations for
three fold degenerate Hubbard model, used to get the self--energy for Ti~$%
t_{2g}$ bands, produce a Mott--Hubbard gap equal to 2.8~eV but once upfolded
into the LDA Hamiltonian one needs to take into account La~$5d$ states in
the vicinity of the Fermi level. The gap between the lower Hubbard band and
La~$5d$ bands is indeed the charge transfer gap and it is equal to
0.2-0.5~eV for the undoped compound. Optical transitions from the lower
Hubbard band to La~$5d$ give the main contribution to the optical
conductivity in pure LaTiO$_{3}$.

\begin{figure}[h]
\epsfig{file=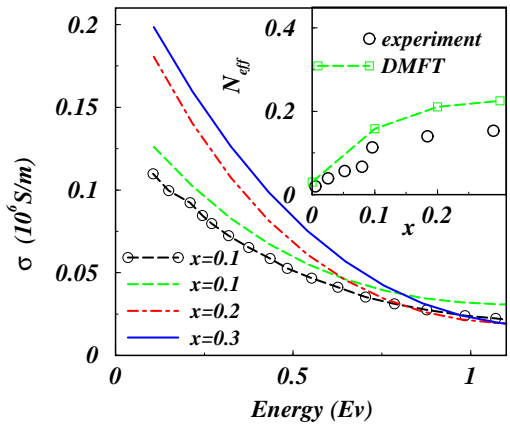,width=2.8in}\\[-0.1cm]
\caption{Low frequency behavior of the optical conductivity for La$_{1-x}$Sr$%
_{x}$TiO$_{3}$ at $x=0.1, 0.2, 0.3$ calculated using the LDA+DMFT method.
Experimental results~\protect\cite{OpticsLTO2} are shown by symbols for the
case $x=0.1$. In the inset the effective number of carriers is plotted as a
function of doping. Squares show the results of the LDA+DMFT calculations.
Circles denote the experimental data from Ref.~\protect\cite{OpticsLTO2}.}
\label{fig:One}
\end{figure}

Upon doping, carriers are introduced, and the system exhibits metallic
behavior. Fig.~\ref{fig:One} shows low--frequency part of $\sigma
_{xx}(\omega )$ at dopings $x=$0.1, 0.2, and 0.3. The optical conductivity
exhibits a Drude peak whose strength is increased with doping. The
contribution to $\sigma _{xx}(\omega )$ at these frequencies is due to
transitions from i) the coherent part of the spectrum near the Fermi level
to the upper Hubbard and Lanthanum bands, ii) the transitions from the lower
Hubbard band to the upper Hubbard band and Lanthanum bands and iii)
transitions from the lower Hubbard band to the coherent part of the spectra.
This trend correctly reproduces the optical absorption experiments performed
for La$_{1-x}$Sr$_{x}$TiO$_{3}$~\cite{OpticsLTO}. Comparison of our data
with these measurements is shown in Fig.~\ref{fig:One} where the measured
optical conductivity at the doping level $x=0.1$ is plotted by symbols.
Overall good agreement can be found for the frequency behavior of the
theoretical and experimental curves.

The strength of the Drude peak is only slightly overestimated by the present
theory as well as some residual discrepancy is seen in the region near 1\
eV. We must emphasize that corresponding calculations based on the local
density approximation would completely fail to reproduce the doping behavior
due to the lack of the insulating state of the parent compound LaTiO$_{3}$.
As a result, the LDA predicts a very large Drude peak even for $x=0$, which
remains \emph{little changed} as a function of doping. In view of these
data, the correct trend upon doping captured by the present calculation as
well as proper frequency behavior can be considered as a significant
improvement brought by this realistic DMFT study.

More insight can be gained by comparing the effective number of
carriers participating in the optical transitions which is
defined by $N_{eff}(\omega _{c})=\frac{2m}{\pi
e^{2}}\int_{0}^{\omega _{c}}\sigma (\omega )d\omega ,$ where $m$
is free electron mass and $\omega _{c}$ is the cut--off energy.
Experimental data for $N_{eff}(\omega _{c})$ are available for the
frequency $\omega _{c}=1.1$ ~eV \cite{OpticsLTO2}. They are shown
in the inset to Fig.~\ref{fig:One} where we plot the effective
number of electrons as a function of hole concentration both from
the theory and experiment \cite{OpticsLTO2}. At zero doping the
system is an insulator which gives very small $N_{eff}$ for $x=0$
(this value is non--zero since we took $\omega _{c}$\ larger than
the optical gap of the insulator). Upon doping, increase in
$N_{eff}$ is expected and its values as well as slope
$dN_{eff}/dx$ agree well with experiment.

The main effect introduced by the DMFT\ calculation on the strength of the
optical transitions can be understood by looking at the Drude and interband
contributions separately and comparing them with the corresponding LDA
values. The LDA data give a very large $N_{eff}=1.15$ which by ninety per
cent consists of the Drude contribution. The latter can be found from the
following equation: $N_{eff}^{D}=\frac{2mV}{\pi e^{2}}\frac{\omega _{p}^{2}}{%
8}$, where plasma frequency $\omega _{p}=4.87$~eV is obtained from
LDA calculations. This result is not surprising since in LDA the
$t_{2g}$ states crossing the Fermi level are filled with one
electron which gives an estimation for the effective number of
electrons participating in optical transitions at this frequency
range. Thus, due to proximity to the insulator the DMFT suppresses
ninety per cent of the Drude part accounted for incorrectly by
the metallic LDA spectrum.

\begin{figure}[h]
\epsfig{file=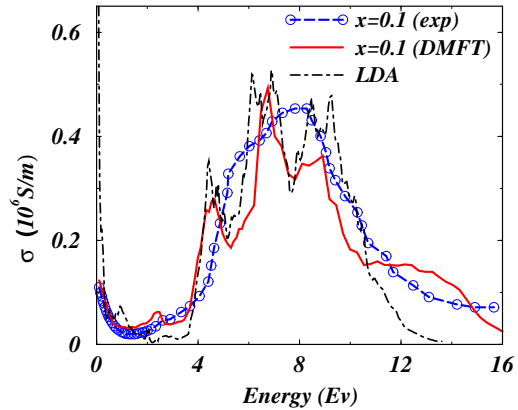,width=2.8in}\\[-0.4cm]
\caption{Calculated using the DMFT optical conductivity spectrum for La$_{x}$%
Ti$_{1-x}$O$_{3}$ with $x=0.1$ at large frequency interval (solid line) as
compared with the experimental data (dashed line with symbols). The results
of the LDA calculations are shown by dashed line.}
\label{fig:opthigh}
\end{figure}

Now we discuss optical conductivity for the frequency interval
from 0 to 16~eV. Fig.~\ref{fig:opthigh} shows $\sigma
_{xx}(\omega )$ at doping $x=0.1$ where we compare our DMFT\ and
LDA\ calculations with the measurements in
Ref.~\onlinecite{OpticsLTO}. Sharp increase in optical
conductivity is seen at $\omega \sim 4$~eV. This can be
attributed to the transitions from the oxygen $p$--band into
unoccupied $d$-states of Ti. The main peak of optical transitions
is located between 5 and 10~eV which is predicted by both DMFT\
calculation (solid line) and the LDA (dashed line). It is
compared well with the measured spectrum (dashed line with
symbols). Since the self--energy corrections modify only the
states near the Fermi level, we do not expect DMFT spectrum to be
essentially different from the LDA one in this frequency range.
Overall, the agreement at high frequencies is quite good which
demonstrates reliability of the present method.

\begin{figure}[h]
\centering\epsfig{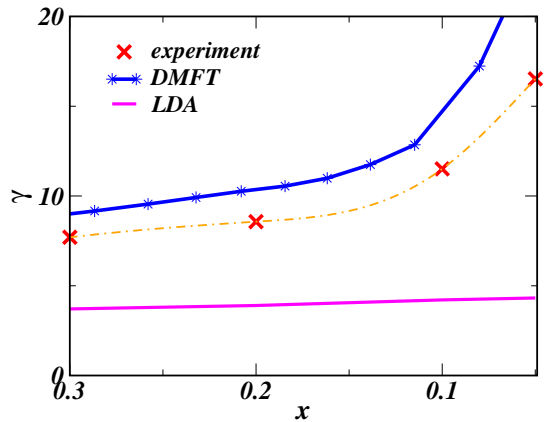}\\[-0.1cm]
\caption{Comparison of the linear coefficient of specific heat, $\protect%
\gamma $, as a function of doping obtained from DMFT (solid line with stars)
and LDA (solid line) calculations against experimental result \protect\cite%
{SpecificHeatExp}. Experimental points are given by cross symbols and
dot--dashed line is used as a guide for eye. }
\label{fig:gamma_qmc}
\end{figure}

As an additional check of the DMFT calculation, we have extracted the values
of the linear specific heat coefficient $\gamma $ as a function of doping.
Our comparisons with the experiment \cite{SpecificHeatExp} are given in
Fig.~ \ref{fig:gamma_qmc}. For example, at $x=0.1$, experimental $\gamma =11%
\frac{ mJ}{molK^{2}}$ while DMFT produces $\gamma $ equal to 14~$\frac{mJ}{%
molK^{2}} .$ Note that the LDA value here is only about 4~$\frac{mJ}{molK^{2}%
}$. Since DMFT renormalizes the density of states at the Fermi level, $%
\gamma $ obtained by this theory clearly indicates the importance of band
narrowing introduced by correlations.

\section{Conclusion}

\label{sec:conclusion}

In conclusion, we have shown how the optical properties of a realistic
strongly correlated system can be computed using recently developed DMFT
based electronic structure method. We have developed a numerically tractable
scheme which is reduced to evaluating dipole matrix elements as well as
integrating in momentum and frequency spaces similar to the methods
developed for non--interacting systems. As an application, we have studied
the optical conductivity of La$_{1-x}$Sr$_{x}$TiO$_{3}$ and found its
correct dependence as a function of frequency and doping in comparison to
the experiment. Our results significantly advance studies based on static
mean--field approximations such as LDA.

The framework that we presented should be a good starting point
for including vertex corrections. Local vertex corrections can be
evaluated within DMFT \cite{DMFTreview} while non--local ones can
be incorporated by extending the calculations of Ref.~\cite{ping}
to the optical conductivity. This is analogous to how LDA spectra
can be improved via the GW method \cite{gwreview}.

\section{Acknowledgments}

This work was supported by the NSF grant DMR-0096462. The authors are
indebted to A. I. Lichtenstein for valuable discussions. We also acknowledge
a warm hospitality extended to three of us (V.O., G.K. and S.S.) during our
stay at Kavli Institute for Theoretical Physics during the workshop
``Realistic Theories of Correlated Electron Materials" where part of this
work has been carried out under Grant No. PHY99-07949.

\appendix

\section{Computation of the self-energy, DMFT downfolding and upfolding}

\label{sec:app_se}

The approach described in Section \ref{sec:method} requires evaluation of
the self--energy operator in Eq.~(\ref{Dyson}) using the LDA+DMFT method
\cite{Anisimov:1997}. The latter exploits the locality of the self--energy
in some orbital space, and the restriction of the Coulomb interaction to a
limited set of localized (or heavy) orbitals to be denoted by $h$. The rest
of the orbitals are taken to be uncorrelated (light) and are denoted by $l$.

Notice that the locality of the self--energy is a basis dependent
statement. Under a change of the basis the Kohn--Sham Hamiltonian,
$H_{\mathbf{k}}$, is transformed into
$U_{\mathbf{k}}HU_{\mathbf{k}}^{\dagger }$, with $U_{\mathbf{k}}$
being a unitary transformation. The self--energy transforms like
the Hamiltonian, however, if $\Sigma (\omega )$ is momentum
independent in one basis, then in the new basis $\Sigma ^{\prime
}=U_{\mathbf{k}}\Sigma (\omega )U_{\mathbf{k}}^{\dagger }$ in
general becomes momentum dependent. Hence, we need to work in a
very localized basis, such as the non--orthogonal LMTO's, where
the DMFT approximation is most justified.

Introduction of a basis set allows the partition of the double--counting
subtracted Kohn--Sham Hamiltonian $H_{hh}^{0}=H_{hh}-\Sigma _{dc}$ and of
the Green's function into the light and heavy blocks:
\begin{eqnarray}
&&G(\mathbf{k},\omega )=\left[ (\omega +\mu )\left( {\begin{array}{*{20}c} {
O_{hh} } & { O_{hl} } \\ { O_{lh} } & { O_{ll} } \\ \end{array}}\right) _{%
\mathbf{k}}\right.  \label{eq:Hilbert_real} \\
&&\left. -\left( {\begin{array}{*{20}c} { H_{hh}^0 } & { H_{hl}^0 } \\ {
H_{lh}^0 } & { H_{ll} }^0 \\ \end{array}}\right) _{\mathbf{k}}-\left( {%
\begin{array}{*{20}c} \Sigma_{hh}(\omega) & 0 \\ 0 & 0 \\ \end{array}}%
\right) _{{}}\right] ^{-1},  \notag
\end{eqnarray}%
where $[...]^{-1}$ means matrix inversion, $\mu $ is the chemical
potential and $O$ is the overlap matrix. Given that the
self--energy is local, it can be obtained from the Anderson
impurity model
\begin{eqnarray}
S_{imp} &=&\sum_{\alpha \alpha ^{\prime },\tau \tau ^{\prime }}c_{\alpha
}^{+}(\tau ){\mathcal{G}_{0}}_{\alpha \alpha ^{\prime }}^{-1}(\tau ,\tau
^{\prime })c_{\alpha ^{\prime }}(\tau ^{\prime })  \label{eq:action} \\
&+&\sum_{\alpha \beta \gamma \delta \text{,}\tau }\frac{U_{\alpha \beta
\delta \gamma }}{2}c_{\alpha }^{+}(\tau )c_{\beta }^{+}(\tau )c_{\gamma
}(\tau )c_{\delta }(\tau ),  \notag
\end{eqnarray}%
where ${\mathcal{G}_{0}}$ is the bath Green's function which obeys the
self--consistency condition~\cite{Anisimov:1997} generalized to
non--orthogonal basis set:
\begin{equation}
{\mathcal{G}}_{0}^{-1}(\omega )=\left( \sum_{\mathbf{k}}{1}\frac{1}{{(\omega
+\mu )}O-H^{0}(\mathbf{k})-\Sigma (\omega )}\right) _{hh}^{-1}+\Sigma
_{hh}(\omega ).  \label{eq:weiss}
\end{equation}

When a group of bands is well separated from the others it is
possible to recast the previous self--consistency condition
\textit{at low frequencies} in a form which resembles the DMFT
equations derived from a Hamiltonian involving the $h$ degrees of
freedom only. In the one--electron approach it goes under the
name downfolding ~\cite{Andersen:2000}.

Performing standard matrix manipulations and a low--frequency
expansion with linear accuracy in $\omega$ (which is justified
for low--energy calculations provided the separation of energy
scales between the band near the Fermi level and the rest) we
rewrite the heavy block of the Green's function as:

\begin{equation}
G_{hh}(\mathbf{k},\omega )=\left[ Z_{\mathbf{k}}^{-1}\omega -\widetilde{H}(%
\mathbf{k})-\Sigma _{hh}\right] ^{-1},  \label{eq:dmft_downf}
\end{equation}%
where renormalization amplitude $Z_{\mathbf{k}}$ and effective Hamiltonian
are given by
\begin{eqnarray}
Z_{\mathbf{k}}^{-1} &=&O_{hh}+K_{hl}K_{ll}^{-1}O_{ll}K_{ll}^{-1}K_{lh}
\notag \\
&&-O_{hl}K_{ll}^{-1}K_{lh}-K_{hl}K_{ll}^{-1}O_{lh},  \notag \\
\widetilde{H}(\mathbf{k}) &=&H_{hh}^{0}-K_{hl}K_{ll}^{-1}K_{lh},  \notag \\
K_{\gamma } &=&H_{\gamma }^{0}-\mu O_{\gamma }.  \label{eq:h_eff}
\end{eqnarray}%
Here $\gamma $ stands for a pair of indices $l$ or $h$. Finally
we perform a unitary transformation $S$ in the heavy block, so as
to work in a nearly orthogonal basis in the $h$-sector:
\begin{equation}
S^{\dagger }{[\sum_{\mathbf{k}}{Z_{\mathbf{k}}}]}^{-1}S=1.
\end{equation}%
Applying this transformation to Eq. (\ref{eq:h_eff}) we arrive to
the local Green's function in the new basis
\begin{equation}
G_{hh}(\omega )=\sum_{\mathbf{k}}\left[ (\omega +\mu )O_{eff}(\mathbf{k}%
)-H_{eff}(\mathbf{k})-\Sigma (\omega )\right] ^{-1},  \label{eq:RD}
\end{equation}%
and to a new DMFT self--consistency condition:
\begin{equation}
\mathcal{G}_{0hh}^{-1}(\omega )=G_{hh}^{-1}+\Sigma (\omega ).
\label{eq:ghh1}
\end{equation}%
This set of equations has clearly the form of the DMFT equations of a model
involving heavy electrons only, with a Hamiltonian and an overlap matrix:
\begin{eqnarray}
O_{eff}(\mathbf{k}) &=&S^{\dagger }Z_{\mathbf{k}}^{-1}S,  \label{eq:sig_upf}
\\
H_{eff}(\mathbf{k}) &=&S^{\dagger }\widetilde{H}(\mathbf{k})S+\mu O_{eff}(%
\mathbf{k}).
\end{eqnarray}

The self--energy $\Sigma $ is still computed from the Anderson
impurity model, but the Coulomb interaction of this model is
renormalized to a smaller effective interaction $U_{eff}$ matrix
\begin{equation}
U_{eff,\alpha ^{\prime }\beta ^{\prime }\gamma ^{\prime }\delta ^{\prime
}}^{\prime }=\sum_{_{\alpha \beta \gamma \delta }}[\sqrt{Z}]_{\alpha
^{\prime }\alpha }[\sqrt{Z}]_{\beta ^{\prime }\beta }[\sqrt{Z}]_{\gamma
^{\prime }\gamma }[\sqrt{Z}]_{\delta ^{\prime }\delta }U_{\alpha \beta
\gamma \delta }.
\end{equation}

Until now the discussion is general, and applies to any system
where there is a set of bands well separated from the rest.
Further simplifications are possible, if we assume that the
system has cubic symmetry and that the overlap $O_{eff}$ is the
unit matrix. For $d$-electrons, cubic symmetry makes the
self--energy and local Green's function diagonal. In this case the
momentum sum in Eq.~(\ref{eq:RD}) can be replaced by the integral
over energy. The local Green's function can be calculated as a
Hilbert transformation
\begin{equation}
G(\omega )=\int_{-\infty }^{+\infty }d\varepsilon \,\frac{D(\varepsilon )}{%
\omega +\mu -\Sigma (\omega )-\varepsilon }.  \label{eq:Hilbert}
\end{equation}%
Here, $D(\varepsilon )$ is the density of states of the the reduced
Hamiltonian $H_{eff}(\mathbf{k})$. Notice that the cubic symmetry keeps $%
U_{eff}$ diagonal if the bare Coulomb matrix $U$ has that property.

Upfolding is a procedure which is \textquotedblleft inverse" to the
downfolding described above. One simply converts the self--energy $\Sigma $
obtained from the DMFT calculation into the block self--energy $\Sigma
_{hh}=S\Sigma S^{\dagger }$, which is to be inserted to the original LDA
Hamiltonian, in order to compute the local Green's function $G(\omega)$.

In general, the downfolded density of states ${D(\varepsilon )}$
obtained from $H_{eff}$ has a non--zero first energy moment and
depends in a non--linear way on the value of the double counting
correction, as well as on the chemical potential which enters the
formulation of the original problem containing all electronic
bands. Furthermore, the value of the chemical potential in the
LDA+DMFT calculations does not need to be the same as the LDA
value.

The reduction of the self--consistent LDA+DMFT equations to the form
described by Eq.~(\ref{eq:Hilbert}) with $D(\varepsilon )$ being the partial
LDA density of states of the heavy orbitals was suggested and used in Ref.~%
\cite{Held:2001}. Unfortunately, this partial density of states
contains weight at high energies, and if this is omitted, the
normalization condition is violated. The derivation presented in
this Appendix eliminates these difficulties, and instead suggests
an alternative procedure in which we first carry out a
tight--binding fit of the LDA bands (downfolding) near the Fermi
level, and then use it to estimate $D(\varepsilon ).$ Our
derivation also indicates how one goes back (i.e. upfolds the
self--energy) to the all--orbital Hamiltonian. In our
calculations using the downfolded equations $\mu $ was adjusted
to get the correct density of $d$--electrons. In the upfolded
Green's function $\mu $ was taken to be the LDA chemical
potential, and ${\Sigma }_{dc}$ was deduced from a constant shift
of the heavy orbitals by obtaining the total number of electrons
from the integral of the spectral function
\begin{equation*}
A(\omega )=-\frac{1}{\pi }\mathrm{Im}\sum\limits_{\mathbf{k}%
}{}\sum\limits_{\alpha \beta }{}G_{\alpha \beta }(\mathbf{k},\omega
)O_{\alpha \beta }^{\mathbf{k}},
\end{equation*}%
multiplied by the Fermi function.

\end{document}